# Automating *Care* by Self-maintainability for Full Laboratory Automation

Koji Ochiai[1,11], Yuya Tahara-Arai[2,3,4,11], Akari Kato[1,4], Kazunari Kaizu[1,5], Hirokazu Kariyazaki[6], Makoto Umeno[7], Koichi Takahashi[1,8,12], Genki N. Kanda[1,4,9,12], Haruka Ozaki[2,4,10,12]

[1] Laboratory for Biologically Inspired Computing, RIKEN Center for Biosystems Dynamics Research. 6-7-1 Minatojima Minamimachi, Chuo-ku, Kobe, Hyogo, 650-0047 Japan
[2] Bioinformatics Laboratory, Institute of Medicine, University of Tsukuba, 1-1-1 Tennodai, Tsukuba, Ibaraki, 305-8577 Japan
[3] Ph.D. Program in Humanics, School of Integrative and Global Majors, University of Tsukuba, 1-1-1 Tennodai, Tsukuba, Ibaraki, 305-8577 Japan
[4] Laboratory Automation Suppliers' Association, 2-1-8 Minatojima Minamimachi, Chuo-ku, Kobe, Hyogo, 650-0047 Japan
[5] Cell Modeling and Simulation Group, Exploratory Research Center on Life and Living Systems, National Institutes of Natural Sciences, 5-1 Higashiyama, Myodaiji, Okazaki, Aichi, 444-8787 Japan
[6] Robotics Division, YASKAWA Electric Corporation, 2-1 Kurosaki Shiroishi, Yahatanishi-ku, Kitakyushu, Fukuoka, 806-0004 Japan
[7] BAIKEIDO LLC, 3-8-1 Asano, Kokurakita-ku, Kitakyushu, Fukuoka, 802-0001 Japan
[8] Graduate School of Media and Governance, Keio University. 5322 Endo, Fujisawa-shi, Kanagawa 252-0882 Japan
[9] Medical Research Laboratory, Institute of Integrated Research, Institute of Science Tokyo, 1-5-45 Yushima, Bunkyo-ku, Tokyo, 113-8519 Japan
[10] Center for Artificial Intelligence Research, University of Tsukuba, 1-1-1 Tennodai, Tsukuba, Ibaraki, 305-8577 Japan
[11] First authors (equal contribution)
[12] Contact authors (equal contribution)

**Contact:**

| | |
|---|---|
| Koichi Takahashi, Ph.D. | ktakahashi@riken.jp |
| Genki N. Kanda, Ph.D., PMP | genki.kanda@riken.jp |
| Haruka Ozaki, Ph.D. | haruka.ozaki@md.tsukuba.ac.jp |

## Abstract

The automation of experiments in life sciences and chemistry has significantly advanced with the development of various instruments and AI technologies. However, achieving full laboratory automation, where experiments conceived by scientists are seamlessly executed in automated laboratories, remains a challenge. We identify the lack of automation in planning and operational tasks—critical human-managed processes collectively termed "care"—as a major barrier. Automating *care* is the key enabler for full laboratory automation. To address this, we propose the concept of self-maintainability (SeM): the ability of a laboratory system to autonomously adapt to internal and external disturbances, maintaining operational readiness akin to living cells. A SeM-enabled laboratory features autonomous recognition of its state, dynamic resource and information management, and adaptive responses to unexpected conditions. This shifts the planning and execution of experimental workflows, including scheduling and reagent allocation, from humans to the system. We present a conceptual framework for implementing SeM-enabled laboratories, comprising three modules—Requirement manager, Labware manager, and Device manager—and a Central manager. SeM not only enables scientists to execute envisioned experiments seamlessly but also provides developers with a design concept that drives the technological innovations needed for full automation.

# Main

## 1. Automating *care* for full laboratory automation

Science is an activity to extend the wisdom of humankind, and the automation of science is the ultimate way to accelerate this activity [1–3]. In the fields of life sciences and chemistry, significant advancements have been made in automating experimental operations, exemplified by the widespread use of automated pipetting machines and laboratory robots [4–13]. However, automating experimental operations alone is insufficient to fully realize the potential of automated scientific discovery. Research inherently involves iterative processes, such as optimizing reaction conditions or exploring parameter spaces, which require adaptive and continuous decision-making. To address this need, the concept of closed-loop experimentation, or self-driving labs (SDLs), have emerged [14], which combines automated experiments with data-driven decision-making powered by AI. These integrated systems have demonstrated remarkable success in various fields, including photocatalysis, microbial genetics, and regenerative medicine [15–20].

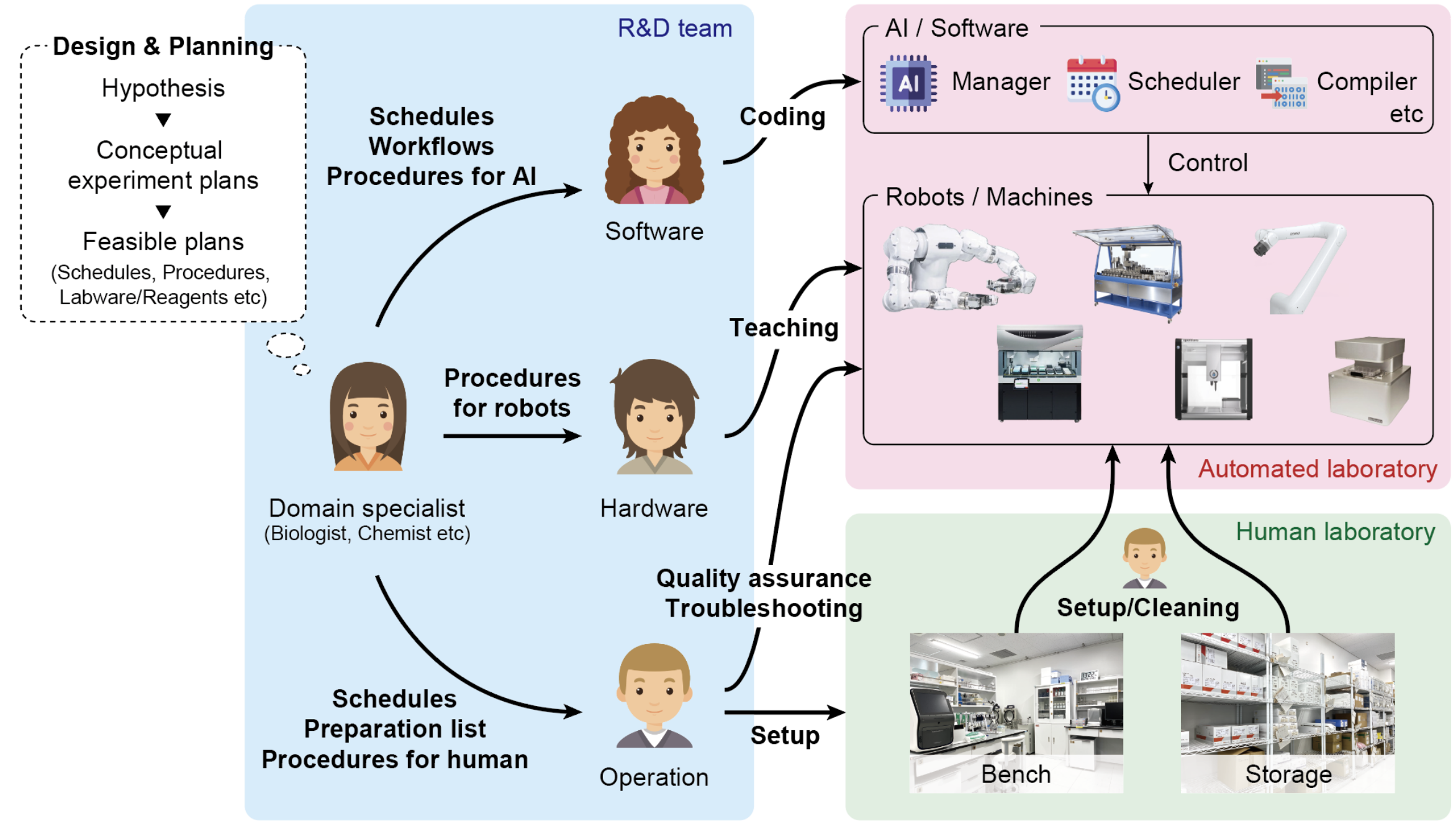

**Figure 1: Human efforts in a conventional laboratory automation workflow (as is)**
This schematic illustrates how human specialists and automated systems collaborate to carry out experiments in biology or chemistry. Bold letters represent human efforts (*care*) in conventional laboratory automation. The R&D team (blue area)—comprising domain specialists (e.g., biologists, chemists), hardware engineers, software engineers, and operators—collectively designs, plans, and executes experiments, creates AI-compatible workflows, and develops robotic procedures. The automated laboratory (pink area) is controlled by AI/Software and houses robots and other automated machines responsible for executing tasks such as sample handling and manipulation. However, human operators remain essential for quality assurance and troubleshooting—for instance, addressing issues like pipetting errors or bubbles—along with setting up and cleaning the lab environment. The human laboratory and storage (green area) require manual tasks such as refilling consumables and preparing specialized equipment. In many cases, data collection also relies on human intervention, depending on the type of observation or instrumentation. While automation can streamline experiments, human care and oversight are critical for ensuring accuracy, maintaining equipment, and resolving unexpected problems. The instruments shown in this figure are examples and represent only part of the available devices: Maholo LabDroid (Robotic Biology Institute Inc.), Fluent (TECAN), Chemspeed (Chemspeed Technologies Inc.), OT-2 (Opentrons Labworks, Inc.), Cobotta Pro (Denso Wave Incorporated), BioStudio-T (Nikon Corporation).

While the integration of robotic automated experiments and AI has enabled the automation of iterative optimization processes, these systems still rely heavily on human involvement for what we propose to define as "care." *Care* refers to the human-managed tasks required to maintain and support automated systems, encompassing a range of essential responsibilities at both the planning and operational steps of automated laboratories (**Figure 1**). In the planning step, care includes translating experimental objectives into machine-readable workflows and scheduling resources such as equipment and reagents. In the operational step, it involves preparing and cleaning up experiments, restocking supplies, monitoring system states, and addressing errors or unexpected conditions. For example, humans must ensure compatibility between workflows and equipment limitations, such as splitting a 200 μL aliquot into two 100 μL portions to align with equipment constraints.. Additionally, tasks like verifying equipment availability and adapting workflows to real-world constraints remain outside the scope of current automation capabilities, leaving these responsibilities as a bottleneck in achieving fully autonomous scientific discovery.

*Care* is indispensable for making experimental protocols feasible under laboratory conditions, bridging the gap between conceptual designs and their practical execution. Usually, experimental protocols are abstract and must be adapted to the physical constraints and conditions of the laboratory: Even with the same conceptual protocol, the operational protocol can differ based on the laboratory environment and setup, due to variations in equipment, space, or resources. Such adjustment processes requires intervention of human managers and operators, primarily because (1) most of the conventional laboratory automation systems lack the ability to recognize their environments and adapt to them, (2) these systems are designed to operate within predefined constraints, relying on user-provided inputs such as initial setups and experimental conditions, (3) the dynamic nature of laboratory environments, where conditions can shift due to concurrent experiments or resource usage, necessitates manual intervention to prepare the setup immediately before execution.

This reliance on human *care* constrains the capabilities of automated laboratories, limiting their application primarily to repetitive or screening-based experiments including closed-loop experimentation. In the planning step, the dependency on human *care* restricts the range and complexity of experiments that can be executed. For example, when introducing new experiments, reagents, and labware, robotic protocols are iteratively refined through trial and error until human operators ensure that workflows are feasible, avoiding flexible modifications in experimental conditions, addition of newly performed experimental protocols, and parallel execution of multiple experiments. In the operational step, the dependency on human *care* (e.g. handling errors) limits the number of experiments that can be executed simultaneously, creating a bottleneck in throughput. As a result, the full potential of automation remains unrealized, with current systems unable to perform the adaptive and flexible processes required for more diverse and complex scientific discovery.

In an ideal automated laboratory, experiments are conducted entirely without human intervention. To achieve this, not only experimental operations but also the tasks we define as *care* must be fully automated. This includes addressing dynamic and unforeseen changes in the laboratory environment, ensuring that experimental processes can continue progressing even in the face of unexpected situations. Such a laboratory would require a new design paradigm in which the system itself autonomously acquires information, makes adaptive decisions, and manages its environment. This approach moves beyond the current reliance on humans to oversee and validate each step, paving the way for a more flexible and truly autonomous scientific discovery process.

In this vision, the entire laboratory functions as a cohesive system, akin to a single, integrated workstation. Instead of viewing individual devices as isolated units, the laboratory operates as a unified whole. Equipment management, labware tracking, waste disposal, and overall status monitoring are automated and coordinated as part of the system. By transferring these responsibilities—conventionally managed by humans—to the automated laboratory itself, the burden on human operators is eliminated. This enables the system to conduct not only repetitive or screening-based experiments but also highly complex, non-standardized workflows that require adaptability and creativity, fulfilling the broader potential of scientific automation.

Implementing automations of *care* through bottom-up extensional engineering is presumably difficult because there are no limits to what can be done. For instance, if there are multiple locations where a tube containing ethanol could be placed, a mechanism could be created to check each of them. However, such a process would be specific to a particular experiment and would require adjustment and redevelopment each time the type of experiment changes. This approach lacks scalability and adaptability for dynamic laboratory conditions.

## 2. Self-maintainability (SeM) for automating *care*

Here, to address these challenges, we propose a new design concept for ideal automated laboratories called **self-maintainability (SeM)**:

> Self-maintainability (SeM) – the system's ability to maintain itself to the state where it can function effectively in environments subject to material consumption, equipment wear, and potential operational disruptions caused by external disturbances.

Conventional approaches to laboratory automation have primarily focused on repeating a single protocol multiple times (**Figure 2A**). However, an ideal automated laboratory should be capable of automatically conducting a wide variety of daily experiments for multiple users (**Figure 2B**). Such an automated laboratory can be characterized by three essential requirements. First, it must integrate multiple devices that operate in concert (**Figure 2C**). Second, to accommodate diverse demands, the laboratory must flexibly incorporate and remove various elements including samples, labware, reagents, waste, and human operators (**Figure 2D**). Third, it must function amid uncontrollable changes, such as ongoing experimental processes, environmental fluctuations, and human intervention (**Figure 2E**). Historically, humans have provided the *care* required to address these challenges, but this care can be automated by incorporating processes analogous to those in living cells—specifically, “intake,” “discharge,” and “metabolism.” Consequently, the laboratory must be able to maintain its own function autonomously: regardless of resource consumption by ongoing experiments, sudden environmental changes, or unforeseen interruptions, it should continue to accept and execute experiments. This capability is here defined as SeM (**Figure 2F**).

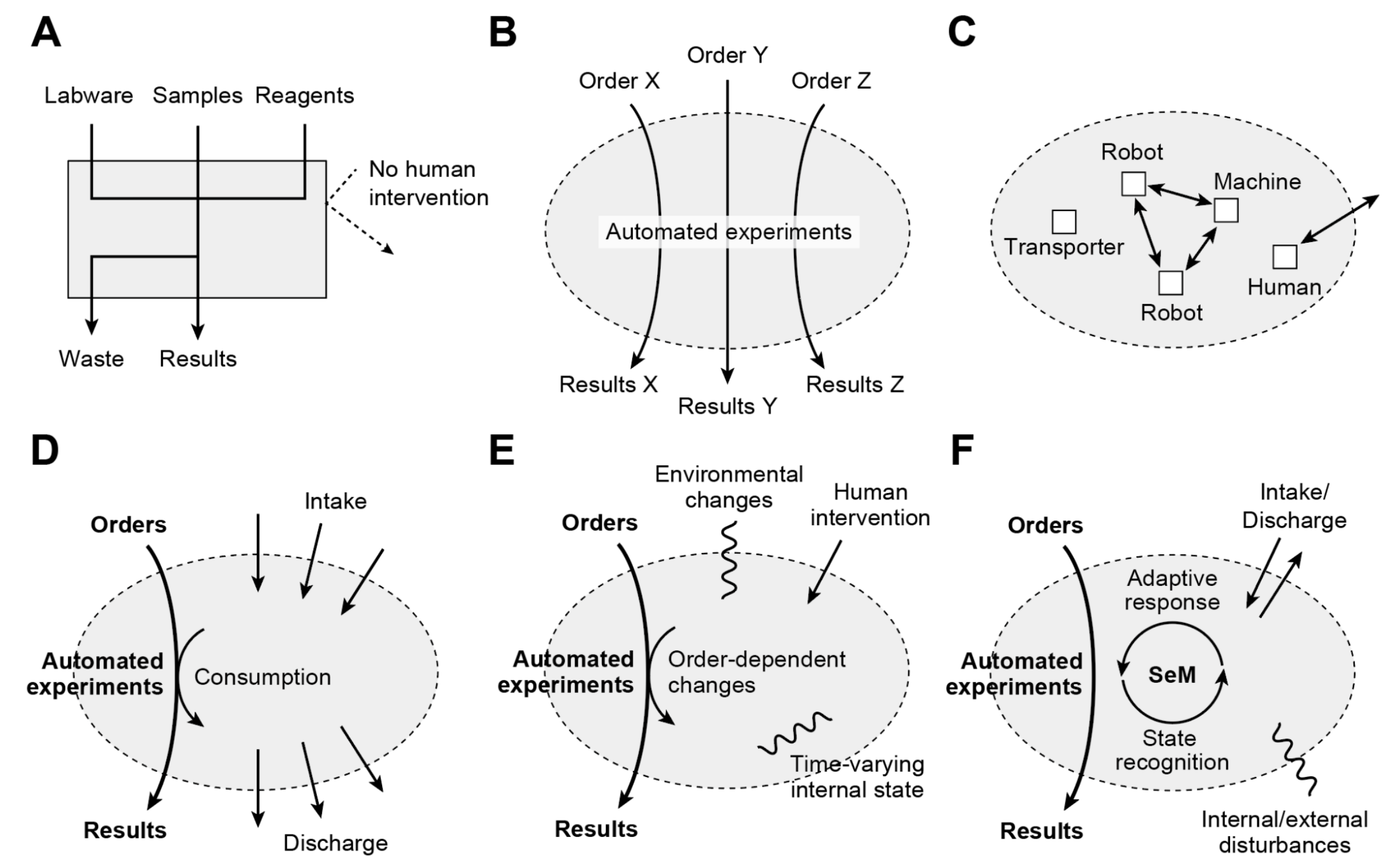

**Figure 2: Conceptual overview of an ideal automated laboratory**
**(A)** Schematic illustration of a conventional automation approach, in which a single protocol is repeated multiple times with minimal human intervention. Labware, samples, and reagents are processed automatically, producing waste and results. The gray rectangle represents the experimental robot.
**(B)** The gray circle denotes the automated laboratory in its ideal state, where multiple orders (X, Y, Z) trigger automated experiments that generate corresponding results.
**(C)** The automated laboratory relies on various components—robots, transporters, machines, and humans—that must cooperate to fulfill experimental demands.
**(D)** In conducting experiments, the laboratory must intake necessary materials and discharge waste, processes that can generally be predicted in advance.
**(E)** The laboratory is subject to various internal and external disturbances, including human intervention. These disturbances cause the situation to change moment by moment, posing obstacles to achieving the ideal state shown in (B).
**(F)** Our newly proposed concept, self-maintainability (SeM), integrates state recognition and adaptive response. This allows the automated laboratory to maintain its function despite internal and external disturbances, much like living cells or organisms do.

SeM provides a holistic framework for designing systems capable of autonomously adapting to and mitigating real-world challenges (**Figure 3**). Existing experiment automation, including most closed-loop experiment systems, has almost no SeM and relies on human *care*, whereas automated laboratories with high SeM can be operated with low burden on humans. In other words, today's automation requires users to think about things that are not important to them (e.g., where to place plates and tube racks on the desk, where to store consumables) each time an experiment is performed. Laboratory's SeM improvements will enable automation of experiments without the need for human effort or time for such matters. For example, tasks currently performed during the planning step, such as translating experimental objectives into machine-readable workflows and scheduling equipment and reagent use, would be autonomously managed by the system. Similarly, operational step tasks like preparing and cleaning up experiments, restocking supplies, monitoring system states, and resolving errors would be handled automatically. SeM not only eliminates repetitive human manual interventions but also enables laboratories to execute more complex and flexible experiments, overcoming the current bottlenecks in scalability and adaptability.

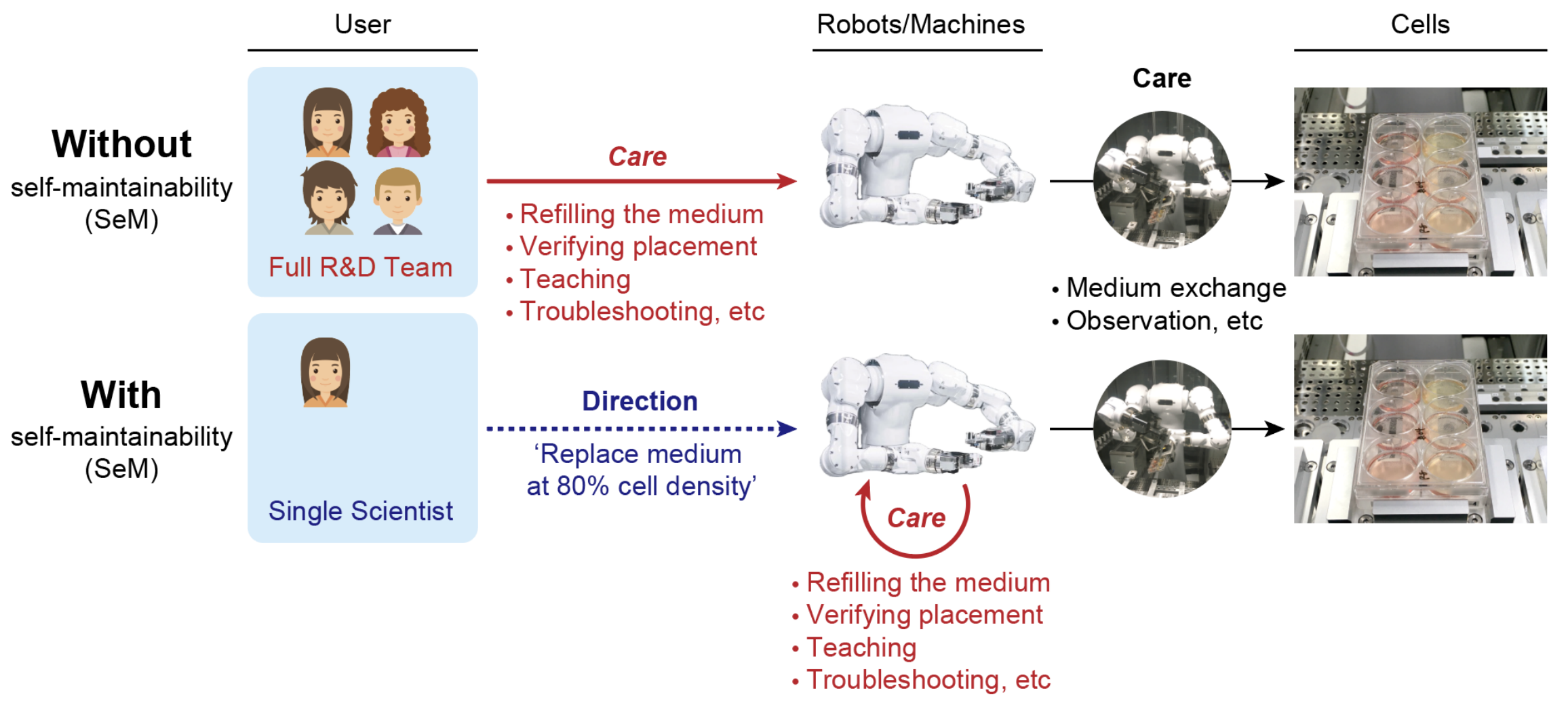

**Figure 3: Self-maintainability (SeM)**
We define self-maintainability (SeM) as the system's ability to maintain itself to the state where it can function effectively in environments subject to material consumption, equipment wear, and potential operational disruptions caused by external disturbances. In the "without self-maintainability" scenario (top), a full R&D team must perform *care* and maintenance tasks (e.g., refilling the medium, verifying placement, teaching, and troubleshooting). By contrast, in the "with self-maintainability" scenario (bottom), a single scientist issues high-level instructions—such as "Replace medium at 80% cell density"—while the system autonomously handles its own *care*.

## 3. SeM-enabled laboratories: Redefining the roles of humans in laboratory automation

Hereafter, an automated laboratory with SeM will be referred to as a SeM-enabled laboratory. Conventional laboratory automation and SeM-enabled laboratories differ significantly in terms of their goal, scope, user assumptions, and handling of information (**Figure 4**). The goal of conventional laboratory automation is to accurately follow the user's instructions, whereas SeM-enabled laboratories aim to fully realize the user's intentions. In terms of scope, conventional systems are limited to single or multiple robots, while SeM-enabled laboratories view the entire laboratory environment as a single, integrated system. Regarding user assumptions, conventional systems treat users as ideal or perfect entities who always provide clear and precise instructions. In contrast, SeM-enabled laboratories acknowledge that users' instructions are inherently incomplete and treat users as sources of uncertainty. In handling information, conventional automation assumes all necessary information is complete and that conditions can be fully calculated or confirmed in advance. On the other hand, SeM-enabled laboratories recognize that information may be incomplete and proactively gather additional data as needed, accounting for internal and external disturbances that might otherwise prevent conditions from being met.

| Conventional laboratory automation (as is) | | SeM-enabled laboratory (to be) |
|---|---|---|
| • Accurately follows the user's instructions | Goal | • Maximally realizes the user's intentions |
| • Single or multiple robots | Scope | • The entire laboratory |
| • Treated as an ideal/perfect entity<br>• Always provides clear and precise instructions | User | • Has its own intentions<br>• Assumes instructions are inherently incomplete<br>• Treated as a source of uncertainty |
| • All provided information is complete<br>• Everything can be fully calculated or confirmed beforehand | Information | • Information may be incomplete; the system adapts by gathering additional data as needed<br>• Internal and external disturbances prevent assuming all conditions are met in advance |

**Figure 4: Key differences between conventional laboratory automation (as is) and a SeM-enabled laboratory (to be)**
Conventional laboratory automation focuses on accurately following user instructions with complete and pre-verified information, typically within the scope of a single or multiple robots. By contrast, a SeM-enabled laboratory treats the entire lab as an integrated system, assumes incomplete information from the outset, and adapts by gathering data as needed. It also takes internal and external disturbances into account, rather than presuming all conditions are satisfied beforehand.

These differences suggest that, in SeM-enabled laboratories, the roles of users and the system are redefined. In existing experimental automation systems, users are responsible for collecting and providing information, and the system determines its actions based on the information provided. Consequently, such systems had to rely on human *care* for various tasks required to complete an experimental order, such as designing and implementing new experimental protocols, responding to changing conditions, and handling errors. In contrast, SeM-enabled laboratories free users from the need to know the internal conditions of the laboratory when making experimental orders. Instead, the SeM-enabled laboratory autonomously collects the necessary information to fulfill the order—without waiting for user instructions—, recognizes the current state, adapts its responses accordingly, and executes automated experiments.

Similarly, tasks related to higher-level experiment management, such as adjusting schedules or adding new experiments, which were previously the user's responsibility, are now handled by the laboratory itself in SeM-enabled laboratory. For instance, in the case of a schedule change, if one of the sample plates is dropped, the user may be asked whether to cancel the experiment or to continue with the remaining plates. Depending on the user's decision, if the experiment is aborted, the laboratory will clean up and discard the samples. If it is continued, unnecessary steps in the protocol (i.e. the subsequent planned experimental steps for the discarded sample plate) will be omitted to adapt to the situation. Additionally, for new experimental orders, the laboratory autonomously manages and adjusts the schedule. For example, it can use available device time to initiate another cell culture or reallocate operations to accommodate a sudden visitor.

## 4. SeM-enabled laboratories implementation example

There are several possible approaches to implement SeM-enabled laboratories, but we believe it is important to enable an automated laboratory to "the ability to recognize the conditions in the laboratory and adapt its behavior accordingly." Existing experiment automation systems require users to perform *care* to conduct an experiment, because they recognize or assume conditions and decide on actions. By providing the ability to automated laboratories, we expect to eliminate the causes of *care* and comprehensively remove the burden from the users.

To realize this ability, it is necessary to redesign the devices and its control system for automated laboratories. First, a mechanism is needed to actively gather additional information as necessary, such as details of experimental conditions, sample priorities, or deadlines, because users are assumed to be unable to fully convey their requirements when making an experimental order. Second, a system that actively collects information about the laboratory's state through sensors is essential. Users often lack complete information about the automated laboratory, such as available equipment, labware, or ongoing experiments. Moreover, the laboratory's state can fluctuate constantly due to internal and external disturbances. Third, it is required to flexibly control the automated experiment systems and robots based on the recognized states, demanding the robot controllers to respond to on-demand motion commands rather than executing predefined actions such as teaching and playback. Fourth, while conventional systems treat sensor data and human input separately, a software architecture is required to handle both sources as they jointly influence experimental changes. Fifth, all information in the laboratory—from users to devices to labware—must always be available, even during experiment execution, to ensure the system can adapt to constantly changing states and accommodate new experimental orders or schedule changes.

Hereafter, we will consider the architecture to realize the SeM-enabled laboratory in more detail. We believe that three modules (Requirement manager, Labware manager, and Device manager) and a Central manager that controls them are necessary to realize the SeM-enabled laboratory (**Figure 5**).

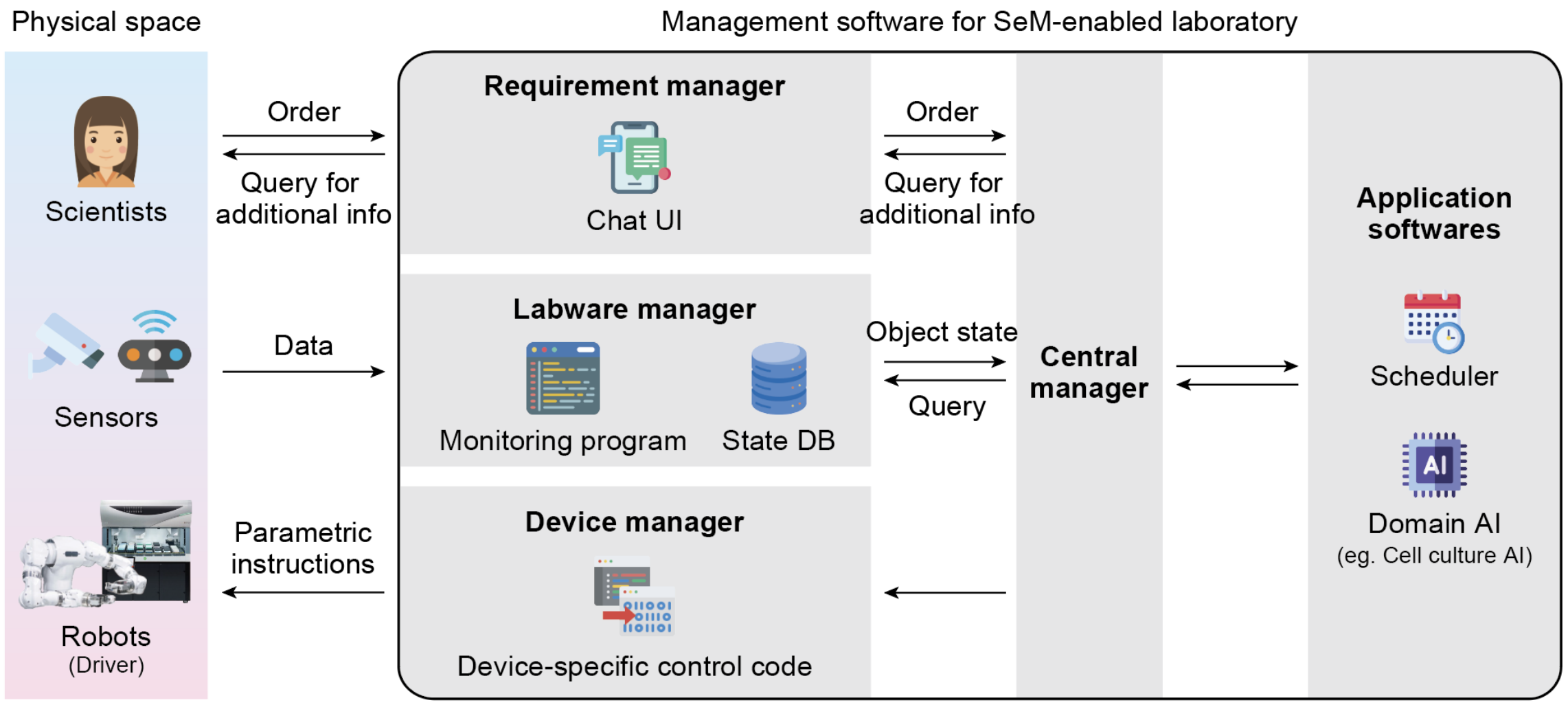

**Figure 5: An example architecture of management software for SeM-enabled laboratory**
A laboratory with SeM is referred to as a SeM-enabled laboratory. The system is composed of the Central manager and three key modules: the Requirement manager, the Device manager, and the Labware manager.

The Requirement manager is a module that communicates interactively with the user to obtain and manage the requirements of an experimental order. To conduct a new experiment in conventional experimental automation systems, a user must (1) create a new protocol with detailed specifications, (2) verify that the protocol achieves the expected behavior using empty labware when necessary, and (3) confirm that the required conditions are met at the start of the protocol before initiating it. In a SeM-enabled laboratory, on the other hand, the SeM-enabled laboratory autonomously plans and determines the details of the experiment with reference to the state of the laboratory. Because it is not assumed that users can state all the information required to complete the experimental order at once, the Requirement manager proactively collects additional information on the requirements for the orders through an interactive interface, such as chat.

The Labware manager is a module that manages the current status of the laboratory by recognizing the state of the laboratory through sensing to deal with disturbances that create differences between the assumed and actual laboratory. Sensing is essential in physical space where external disturbances cannot be eliminated, making it impossible to predetermine the state of all objects. Even if a part of the experimental devices is isolated from external factors, some parts of the laboratory must remain open to external interactions to facilitate the entry and exit of consumables. Consequently, external disturbances may occur, such as misplaced consumables or variations in consumable shapes due to lot-to-lot variability. Image recognition is considered useful for recognizing conditions in the laboratory. Recent studies such as YOLO [21] and ot2eye [22], which applied YOLO to labware recognition, can be used to realize SeM. Experimental systems also need the ability to estimate the current state while using past observations. For example, even if a tube is now out of the camera's field of view, it should remain at the last observed position, or the amount of reagent should have decreased since it has been dispensed.

The Device manager is the module that translates the actions determined by the Central manager into specific instructions for the robot or device. The Device manager encompasses device-specific programs to control various devices and robots in a SeM-enabled laboratory. Depending on the state recognized by the Labware manager, the Central manager determines the actions required to achieve a desired state. The Device manager must be flexible enough to adapt to such instructions by the Central manager. While conventional solutions such as dedicated application programming interfaces (APIs) can specify locations in an automated workstation using symbolic position specification. However, in a SeM-enabled laboratory, where multiple devices and robots are used, such solutions as they complicate cross-device coordination (e.g., object transfer and liquid handling across equipments) and make it difficult to handle unexpected situations caused by external disturbances (e.g., misplaced consumables). To address this, the Device manager requires parametric instructions that can flexibly specify positions in physical space, enabling coordination across devices and adjustments for unexpected states. Recently, motion generation AI for robots such as RT-2 [23] has been studied, and it is expected that this technology can be applied to SeM-enabled laboratories.

These three modules are controlled from a higher level module, the Central manager, which is intended to be a generative AI like foundation model. The Central manager uses information from users and sensors through the Device manager and the Labware manager to design and proceed experiments through a trial and error process. Trial and error here is not limited to through actuators, but also includes resource management such as allocation of consumables and adjustment of equipment usage time with other experiments. As shown in **Appendix A**, in order to conduct an experiment, it is necessary to determine the experimental design by estimating the necessary resources and repeatedly checking and reserving resources several times. Until now, these tasks have been performed by humans, but with a SeM-enabled laboratory, these tasks are performed by the Central manager.

A SeM-enabled laboratory is realized through the coordinated operation of four modules: the Requirement manager, Labware manager, Device manager, and Central manager. To illustrate the differences between conventional and SeM-enabled laboratories, consider the following scenario. In conventional automated systems equipped with an automated pipetting machine, if a user wants to “add 2 mL of cell culture medium to all wells of this plate,” they must perform the following *care*: (1) check the available time of the pipetting machine, (2) make a reservation, (3) prepare medium-containing tubes, (4) place the tubes on the deck of the pipetting machine together with the plate, and (5) create and execute an operation program that matches the placement of the tubes.

In contrast, in a SeM-enabled laboratory, when a user requests, “Add 2 mL of cell culture medium to all wells of this plate” via the Requirement manager, the Central manager first collaborates with the Labware manager to reserve an automated pipetting machine and medium-containing tubes. Next, the Central manager uses the Device manager to operate a robotic arm to move the plate and tubes to the automated pipetting machine. Finally, the Central manager operates the automated pipetting machine, again via the Device manager, to transfer medium from the tube to the plate. In this way, SeM-enabled laboratories can automate tasks that were previously performed by human operators.

## 5. Conclusion and Discussion

The automation of diverse human tasks in laboratories has long been recognized as essential for stabilizing experimental quality, reducing costs, and improving efficiency. Over the years, various design concepts and implementation strategies have been proposed to advance laboratory automation. These include the development of specialized equipment for automating complex experimental operations and the use of general-purpose robots to handle diverse laboratory tasks [24–26]. To automate sample transfer between instruments, mobile robots and rail-mounted robotic arms have been employed [27,28]. Proposals for standardized levels and performance metrics of laboratory automation have further clarified goals and benchmarks for achieving greater integration and efficiency [14,29,30]. Frameworks for integrating mobile robots with device networks in physical spaces [31] and unified APIs for device control [32] and inter-device communication [33] have been proposed to enhance device coordination. Hierarchical software architectures based on system modeling have also been introduced to control highly automated laboratories [34]. Commercial solutions now provide orchestration software that connects multiple devices with mobile robots [35,36]. Simplifying and automating experimental protocol descriptions have been enabled by programming languages for abstract protocol design [37]. Furthermore, in fields such as cell engineering, the importance of addressing internal and external disturbances in automated laboratories has been emphasized [38]. Innovations include digitization and automation of inventory management to ensure seamless resource availability [39] and advancements in laboratory information management systems to streamline data flow and workflow coordination [40]. Additionally, data representation and exchange schemes have been developed to facilitate efficient communication and interoperability between systems [41–43]. Automating experiment scheduling [44] and enabling iterative optimization by integrating algorithms like Bayesian optimization, which adapt experimental plans based on results, have also contributed to these advancements [14]. The vision of fully automated laboratories, capable of solving global challenges, remains a shared aspiration for humanity [2]. Despite these advancements, fundamental gaps remain in automating critical human tasks to ensure compatibility, resource availability, and workflow feasibility, including planning tasks (e.g., translating objectives into workflows, scheduling resources) and operational tasks (e.g., experimental setup, monitoring, and error handling).

In this paper, we addressed the critical yet often overlooked concept of *care*—the human-managed tasks essential for maintaining and supporting automated systems. By proposing SeM as a key enabler of fully automated laboratories, we redefined the automated laboratory as a holistic, adaptive system capable of managing dynamic environments and responding to unexpected disturbances without human intervention. SeM enables laboratories to internalize tasks conventionally performed by humans—such as resource management, error handling, and workflow adaptation—bridging the gap between human intentions and the practical execution of automated experiments. The concept of SeM-enabled laboratories marks a paradigm shift, treating the entire laboratory as a unified, self-regulating entity. We also explored the technical considerations necessary to redesign devices and their control systems to achieve SeM-enabled laboratories. We believe SeM-enabled laboratories not only eliminate bottlenecks stemming from human dependencies but also extend the capabilities of automation to encompass complex, flexible, and diverse experimental workflows, paving the way for a new era of autonomous scientific discovery.

SeM is also crucial for the application of AI-based scientific research, which has recently attracted attention in the field of machine learning, to the biological and scientific fields. Leu et al. recently developed “AI Scientist” and demonstrated the automation of science in the field of machine learning [45]. The AI Scientist automates the whole processes of machine learning research including knowledge retrieval, hypothesis generation, validation experiments, data analysis, paper writing, and peer review. The machine learning field is a good benchmark for automation in science as an early demonstration because the experimental process is completed inside the computer. On the other hand, applying this kind of framework to experimental science fields such as life science and chemistry presents a unique challenge. It requires enabling AI systems to autonomously execute experimental processes in the physical space without human intervention or *care*. SeM-enabled laboratories are fundamental to overcoming this challenge, providing the infrastructure necessary to realize fully autonomous experimental research.

## Data availability

No primary experimental results, data, software or code have been included in this paper.

## Author contributions

Conceptualization, K.O.; Investigation, all authors; Writing - Original Draft, K.O, A.K., G.N.K., H.O.; Visualization, K.O., K.K., G.N.K.; Supervision, K.T., G.N.K, H.O.; Project administration, K.O., A.K.; Funding acquisition, A.K., K.T., G.N.K, H.O.

## Conflicts of interest

K.O., K.K., and K.T. are inventors on patents and patent applications owned by RIKEN covering self-maintainability. H.K. is an employee of YASKAWA Electric Corporation. M.U. is an executive of BAIKEIDO LLC. All the other authors declare no competing interests.

## Acknowledgments

We thank T. Mitsuyama (AIST) and Y. Tanaka (TEGAMI Robotics) for their insightful discussion in the initial meeting; L. Valeggia, M. R. Licheri, J. Meredith, and Y. Kido (TECAN Group Ltd.) for invaluable discussions. We also thank all the lab members at RIKEN BDR, especially R. Nakanishi for insightful discussion. The icons used in this paper were provided by Freepik, dreamicons and iconixar from Flaticon under the free Flaticon License.

## Funding

This study was supported by grants from the JST-Mirai Program (JPMJMI18G4 and JPMJMI20G7 to K.T. and H.O.), Grant-in-Aid for Early-Career Scientists (JP22K17992 to H. O.), Grant-in-Aid for Transformative Research Areas (A) (JP23H04149 to H.O.), Grant-in-Aid for JSPS Fellows (JP22KJ3148 to A.K.), and Grant-in-Aid for Scientific Research (C) (JP23K11820 to G.N.K.).

# Appendix

## A. Flow of executing a specific protocol in the conventional automated laboratory

1. Define a research query
    1.1. I want to find the key process for inducting retinal pigmented epithelial (RPE) cells from iPS cells
2. Determine the outline
    2.1. Learn the protocol from the collaborative research partner
    2.2. First passages of cells and then differentiate them
    2.3. Use 4 different medium types in one induction of differentiation
    2.4. Sampling before and after changing media types
    2.5. Evaluate by NGS
3. Consider the scope of what the robot will be used (estimate the resources needed without considering resource constraints)
    3.1. Which robot to use?
        3.1.1. Use Company A's laboratory robot
    3.2. Is the necessary equipment connected to the robot?
        3.2.1. Check that the $CO_2$ Incubator is connected to the robot
    3.3. Which scope of work will be performed by staff (do it yourself or outsource)
        3.3.1. If the need for reagent replenishment arises, will a staff do it
        3.3.2. Sampling is done by a staff (the robot just hands the sample to the staff)
        3.3.3. Analysis is outsourced
4. Consider the format of the sample
    4.1. Each plate will be cultured at 3 wells/plate (prepare 3 wells of sample for NGS)
    4.2. Culture each well at 2 mL/well of medium
    4.3. Collect one plate per sampling
5. Consider relative schedules
    5.1. Consider the timescale of the schedule
        5.1.1. Schedule in days this time because of the slow iPS growth rate
    5.2. Day 0
        5.2.1. Before passaging - sampling_1
        5.2.2. Passage
        5.2.3. Start induction of differentiation with medium a
    5.3. Day 1
        5.3.1. 24 hours after starting culture on medium a - sampling_2
        5.3.2. Change medium with medium a
    5.4. Day 2
        5.4.1. Day off
    5.5. Days 3-6
        5.5.1. Change media once a day with medium a: 4 days
    5.6. Day 7
        5.6.1. Before changing to medium b - sampling_3
        5.6.2. Change medium with medium b
    5.7. Day 8
        5.7.1. 24 hours after starting culture on medium b - sampling_4

5.7.2. Change medium with medium b
5.8. Day 9
5.8.1. Day off
5.9. Days 10-13
5.9.1. Change media once a day with medium b: 4 days
5.10. Day 14
5.10.1. Before changing to medium c - sampling_5
5.10.2. Change medium with medium c
5.11. Day 15
5.11.1. 24 hours after starting culture on medium c - sampling_6
5.11.2. Change medium with medium c
5.12. Day 16
5.12.1. Day off
5.13. Days 17-20
5.13.1. Change media once a day with medium c: 4 days
5.14. Day 21
5.14.1. Before changing to medium d - sampling_5
5.14.2. Change medium with medium d
5.15. Day 22
5.15.1. 24 hours after starting culture on medium d - sampling_6
5.15.2. Change medium with medium d
5.16. Day 23
5.16.1. Day off
5.17. Days 24-27
5.17.1. Change media once a day with medium d: 4 days
5.18. Day 28
5.18.1. Last sampling - sampling_9

6. Consumables Estimate
6.1. Amount of medium
6.1.1. Medium a: 0(8), 1(7), 3(7), 4(7), 5(7), 6(7) 258 mL *Number of Days (plates)
6.1.1.1. 2 mL x 3 wells x 8 plates x 1 time = 48 mL
6.1.1.2. 2 mL x 3 wells x 7 plates x 5 times = 210 mL
6.1.2. Medium b: 7(6), 8(5), 10(5), 11(5), 12(5), 13(5) 186 mL
6.1.2.1. 2 mL x 3 wells x 6 plates x 1 time = 36 mL
6.1.2.2. 2 mL x 3 wells x 5 plates x 5 times = 150 mL
6.1.3. Medium c: 14(4), 15(3), 17(3), 18(3), 19(3), 20(3) 114 mL
6.1.3.1. 2 mL x 3 wells x 4 plates x 1 time = 24 mL
6.1.3.2. 2 mL x 3 wells x 3 plates x 5 times = 90 mL
6.1.4. Medium d: 21(2), 22(1), 24(1), 25(1), 26(1), 27(1) 42 mL
6.1.4.1. 2 mL x 3 wells x 2 plates x 1 time = 12 mL
6.1.4.2. 2 mL x 3 wells x 1 plates x 5 times = 30 mL
6.2. Amount of reagent for passage
6.2.1. PBS: 2 mL x 3 well x 8 = 48 mL
6.2.2. Cell dissociation solution: 1.5 mL x 3 well x 8 = 36 mL
6.3. Amount of reagent for cleaning
6.3.1. Aspirator is cleaned by aspirating 70% ethanol once a day
6.3.2. 70% ethanol 20 mL x 29 = 580 mL

6.4. Labware
  6.4.1. 6 well plate x 9 plates
  6.4.2. 50 mL tubes (up to 30 mL per tube)
    6.4.2.1. Medium: 9 + 7 + 4 + 2 = 23 tubes using-up
    6.4.2.2. PBS: 2 tubes using up
    6.4.2.3. Cell dissociation solution: 2 tubes using up
    6.4.2.4. 70% ethanol: 2 tubes reuse (Add 70% ethanol to the same tube for reuse)

7. Checking the necessary information for scheduling
  7.1. Check the reservation status of robots and laboratory equipment.
    7.1.1. Company A's experimental robot No. 2 is available from 2 to 3.5 months from now, so we will use it
    7.1.2. Check that the CO2 incubator connected to the experimental robot can hold more than 9 plates
  7.2. Check the schedules of the staff
    7.2.1. Check the schedules of the technical staff, etc
  7.3. Check the inventory of Labware to be used
    7.3.1. Check that there are enough plates and 50 mL tubes
    7.3.2. Reserve labware in a container for the project
  7.4. If there is any Labware lacking, order it
    7.4.1. Order new reagents for induction of differentiation since
    7.4.2. Confirm quotation and delivery date
  7.5. Confirm how to outsource the analysis
    7.5.1. Confirm that 4 wells of samples are required
  7.6. Securing the necessary budget for implementation
    7.6.1. After checking the budget, it was found that there was relatively room to increase the number of wells
    7.6.2. We will prepare 5 wells

8. Estimate again
  8.1. Consider relative schedules
    8.1.1. No change this time
  8.2. Consumables Estimate
    8.2.1. Amount of medium
      8.2.1.1. Medium a: 0(8), 1(7), 3(7), 4(7), 5(7), 6(7) 430 mL *Number of Days (plates)
        8.2.1.1.1. 2 mL x 5 wells x 8 plates x 1 time = 80 mL
        8.2.1.1.2. 2 mL x 5 wells x 7 plates x 5 times = 350 mL
      8.2.1.2. Medium b: 7(6), 8(5), 10(5), 11(5), 12(5), 13(5) 310 mL
        8.2.1.2.1. 2 mL x 5 wells x 6 plates x 1 time = 60 mL
        8.2.1.2.2. 2 mL x 5 welsl x 5 plates x 5 times = 250 mL
      8.2.1.3. Medium c: 14(4), 15(3), 17(3), 18(3), 19(3), 20(3) 190 mL
        8.2.1.3.1. 2 mL x 5 wells x 4 plates x 1 time = 40 mL
        8.2.1.3.2. 2 mL x 5 wells x 3 plates x 5 times = 150 mL
      8.2.1.4. Medium d: 21(2), 22(1), 24(1), 25(1), 26(1), 27(1) 70 mL
        8.2.1.4.1. 2 mL x 5 wells x 2 plates x 1 time = 20 mL
        8.2.1.4.2. 2 mL x 5 wells x 1 plates x 5 times = 50 mL
    8.2.2. Amount of reagent for passage
      8.2.2.1. PBS: 2 mL x 5 well x 8 = 80 mL

8.2.2.2. Cell dissociation solution: 1.5 mL x 5 well x 8 = 60 mL

8.2.3. Amount of reagent for cleaning

8.2.3.1. Aspirator is cleaned by aspirating 70% ethanol once a day

8.2.3.2. 70% ethanol 20 mL x 29 = 580 mL (No change)

8.2.4. Labware

8.2.4.1. 6 well plate x 9 plates

8.2.4.2. 50 mL tubes (up to 30 mL per tube)

8.2.4.2.1. Medium 16 + 11 + 7 + 3 = 36 tubes using-up

8.2.4.2.2. PBS 3 tubes using up

8.2.4.2.3. Cell dissociation solution 2 tubes using up

8.2.4.2.4. 70% ethanol 2 tubes reuse (Add 70% ethanol to the same tube for reuse)

9. Checking the necessary information for scheduling again

9.1. Check the reservation status of robots and laboratory equipment

9.1.1. Company A's experimental robot No. 2 is available from 2 to 3.5 months from now, so we will use it (No change)

9.1.2. Check that the CO2 incubator connected to the experimental robot can hold more than 9 plates (No change)

9.2. Check the schedules of the staff

9.2.1. Check the schedules of the technical staff, etc (No change)

9.3. Check the inventory of Labware to be used

9.3.1. Secure additional reagents

9.4. If there is any Labware lacking, order it

9.4.1. Establish a quote again with 5 wells

9.4.2. Place an order and manage the delivery date

9.5. Confirm how to outsource the analysis

9.5.1. 4 well samples are required, and 5 well samples are sufficient

9.6. Securing the necessary budget for implementation

9.6.1. Confirm that the budget is sufficient for a 5-well estimate

10. Consider a specific schedule

10.1. Start date and time

10.1.1. Adjust the period during which equipment is secured so that sampling and other human work do not overlap with weekends as much as possible

10.2. Refill date and time

10.3. Sampling time

10.4. Adjust staff schedule if human operation is required

10.4.1. Adjust the schedule so that staff can be on standby at the sampling time

10.5. Notify people using the same robot

11. Think about and register specific actions

11.1. Consider and create initial placement and operation protocols for each of the following: passaging, medium exchange, and 70% ethanol aspiration

11.1.1. Initial arrangement: Initial arrangement of labware (reagents, consumables, tips, etc.) on the experiment automation system

11.1.2. Operating protocol: Operating protocol for the laboratory robot

11.1.3. The medium exchange is created for each target plate

12. Create an operating procedure

13. Run the robot according to the specific schedule
14. Collect samples from the freezer and send them to the outsourcing company for analysis